\documentclass[conference]{IEEEtran}
\IEEEoverridecommandlockouts
\usepackage{algorithm}  
\usepackage{algorithmic}
\usepackage{amsmath,amsfonts}
\usepackage{cases}
\usepackage{mathtools}
\usepackage{amssymb}
\usepackage{bm}
\usepackage{relsize}
\usepackage{array}
\usepackage{textcomp}
\usepackage{float}
\usepackage{stfloats}
\usepackage{url}
\usepackage{verbatim}
\usepackage{graphicx}
\usepackage{subfigure}
\usepackage{cite}
\usepackage{color}
\usepackage{makecell}
\usepackage{etoolbox}
\usepackage{xparse} 
\usepackage{microtype}

\newtheorem{lemma}{\underline{Lemma}}

\def\BibTeX{{\rm B\kern-.05em{\sc i\kern-.025em b}\kern-.08em
	T\kern-.1667em\lower.7ex\hbox{E}\kern-.125emX}}

\allowdisplaybreaks

\begin{document}

\title{Robust Beamforming for MIMO Radar with Imperfect Prior Distribution Information}

\author{
	\IEEEauthorblockN{Yizhuo Wang and Shuowen Zhang}
	\IEEEauthorblockA{	Department of Electrical and Electronic Engineering, The Hong Kong Polytechnic University\\[-2pt]
		E-mails: yizhuo-eee.wang@connect.polyu.hk, shuowen.zhang@polyu.edu.hk}
}

\maketitle
\begin{abstract}
	This paper studies a multiple-input multiple-output (MIMO) radar system for sensing the \emph{unknown} and \emph{random} angular location (angle) of a point target, based on the target-reflected echo signals and known prior distribution information about the target's angle specified by a probability density function (PDF). We consider a challenging yet practical scenario where the knowledge of such PDF is \emph{imperfect}, due to the inaccuracy in PDF acquisition or unpredicted change of target appearance pattern; while the real (actual) PDF is modeled as an unknown perturbed version of the imperfect known PDF bounded by a given uncertainty radius. Such PDF imperfection motivates us to study the \emph{robust transmit beamforming} design to optimize the \emph{worst-case} sensing performance among all possible real PDFs. Since the sensing mean-squared error (MSE) is difficult to be characterized explicitly, we adopt the \emph{worst-case posterior Cram\'{e}r-Rao bound (PCRB)} as the performance metric. We formulate the beamforming optimization problem to minimize the maximum PCRB among all possible real PDFs, which is highly non-trivial since the PCRB has a complex intractable expression over the real PDF, and  there are infinite constraints corresponding to the continuous set of real PDFs bounded by the uncertainty radius. To address these challenges, we derive a tractable quadratic approximation of the PCRB via \emph{second-order Taylor expansion}, and leverage the \emph{S-procedure} to equivalently transform the infinite constraints into a linear matrix inequality, based on which the problem is reformulated into a convex optimization problem solvable with polynomial time complexity. The obtained solution approaches the globally optimal robust beamforming solution as the uncertainty radius decreases. Numerical results validate the effectiveness of our proposed robust beamforming design.
\end{abstract}


\vspace{-1mm}
\section{Introduction}\label{sec_int}

Multiple-input multiple-output (MIMO) radar has garnered substantial research interests due to its ability of boosting the sensing performance via exploitation of waveform diversity \cite{fishler2004mimo}. Unlike phased-array radar, MIMO radar can transmit independent waveforms from multiple antennas, thus offering larger virtual aperture, enhanced spatial resolution, improved parameter identifiability, and greater flexibility in beamforming design \cite{li2008mimo}. To fully harness the degrees-of-freedom (DoFs) provided by MIMO radar, judicious design of the transmit signals (or ``\emph{beamforming}'') is of paramount importance.

Existing literature on MIMO radar has predominantly adopted two sensing performance metrics for beamforming design. The first is the beampattern similarity with a desired pattern \cite{stoica2007probing}, which is tractable but cannot explicitly reflect the mean-squared error (MSE). The second is the Cram\'{e}r-Rao bound (CRB), which provides a lower bound on the MSE of any unbiased estimator \cite{li2007range,jointDesign_CRB1,hua2024mimo}. Despite the ground truth and useful insights provided by CRB-based designs, CRB is a function of the \emph{exact values} of the parameters to be sensed, \textls[0]{which are typically \emph{unknown a priori} in practical sensing scenarios.} \looseness=-1

\begin{figure}[t]
	\centering
	\includegraphics[width=0.4\textwidth]{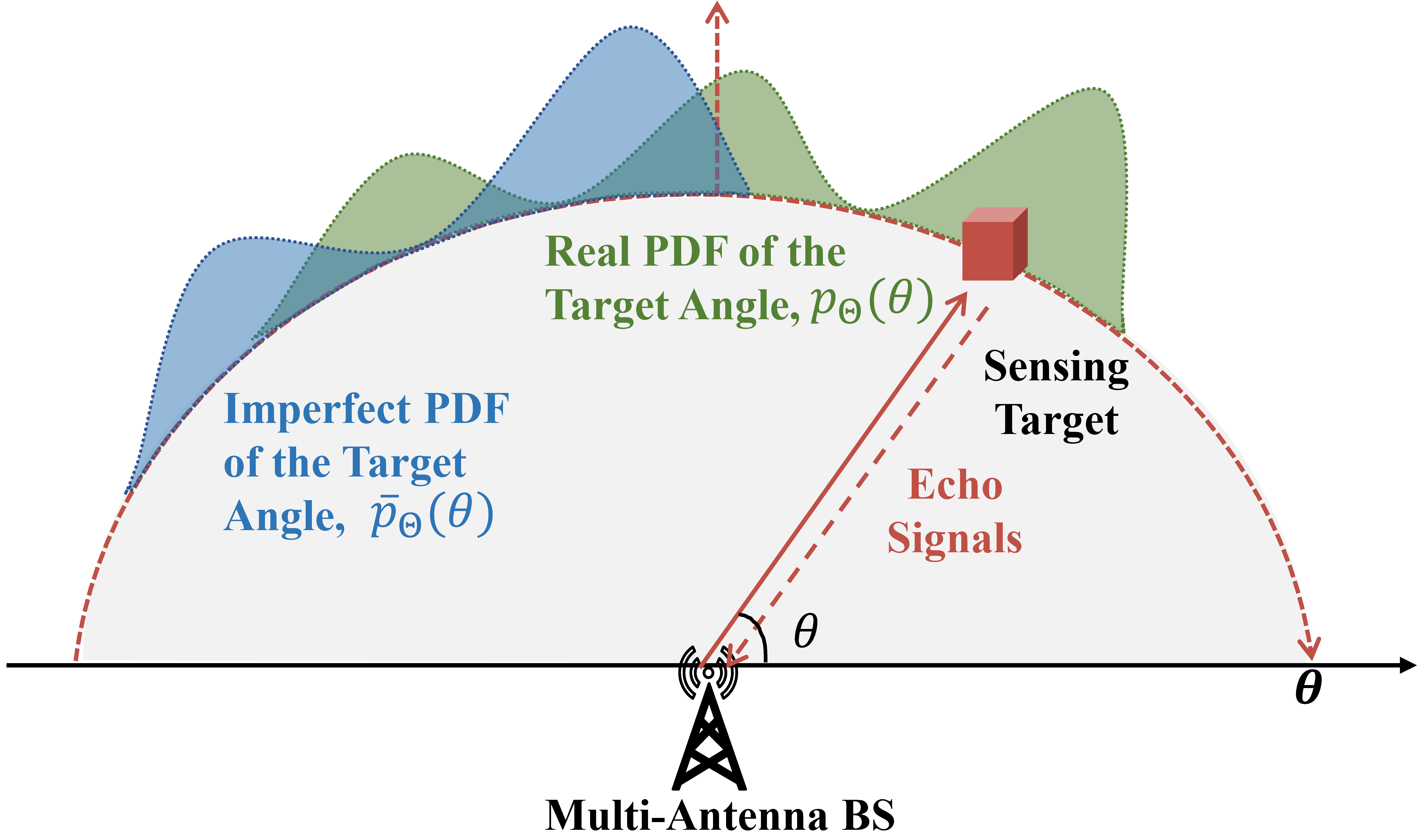}
	\vspace{-3mm}
	\caption{Illustration of a MIMO radar system with imperfect prior information.}
	\label{system model}
\end{figure} 
In practice, the parameters to be sensed are often \emph{unknown} and \emph{random}, while their statistical distribution information can be known \emph{a priori} from historical data or target properties \cite{xu2023radar, xu2024mimo, yao2024gc, hou2023optimal, yao2025optimal, liu2024ris, wang2025hybrid, zheng2025beyond,Yu_Bound2,ghaddar2025active}. The MSE exploiting such prior distribution information is lower bounded by a so-called \emph{posterior Cram\'{e}r-Rao bound (PCRB)} \cite{van1968detection}, which only depends on the probability density function (PDF) of the parameters and is tight in moderate-to-high signal-to-noise ratio (SNR) regimes \cite{van1968detection}. With PCRB as the sensing performance metric, the optimal beamforming design in MIMO radar systems was proved to exhibit a novel ``\emph{probability-dependent power focusing}'' effect \cite{xu2023radar,xu2024mimo,yao2024gc} where high power tends to be concentrated around highly-probable target locations. The line of research on PCRB-based beamforming design was also extended to various integrated sensing and communication (ISAC) systems \cite{hou2023optimal, yao2025optimal, liu2024ris, wang2025hybrid, zheng2025beyond,Yu_Bound2,ghaddar2025active}.

The above studies focused on the ideal assumption that the prior distribution information (i.e., PDF) is \emph{perfectly known}. However, in practice, the knowledge of such information may be \emph{imperfect} due to inaccuracies in PDF acquisition (e.g., from insufficient historical data), unpredicted change of target appearance pattern, etc. If the effect of such imperfect prior distribution information is not taken into account in the beamforming design, the resulted beampattern may yield poor sensing performance under the actual real PDF, e.g., no power is beamed towards the target's actual highly-probable locations when such locations have zero probability densities in the imperfect known PDF, as illustrated in Fig. \ref{system model}.

Motivated by this critical issue, this paper aims to make the first attempt to propose a \emph{robust beamforming} design for MIMO radar with \emph{imperfect prior distribution information}. We focus on a MIMO radar system where a multi-antenna base station (BS) aims to sense the \emph{unknown} and \emph{random} angle of a point target based on the target-reflected echo signals and the known prior PDF of the angle, which is \emph{imperfect}. The real (actual) PDF is modeled as a \emph{perturbed} version of the imperfect known PDF, which lies in an uncertainty set bounded by a given uncertainty radius.\footnote{This model is inspired by the modeling of imperfect channel state information in communication \cite{pascual2005robust,zheng2008robust,robust_xu}. It is worth noting that due to the continuous angle domain of the PDF function (in contrast to the finite dimensions of channel vectors) as well as the complex and intractable sensing performance metric with respect to the PDF, the robust beamforming design for MIMO radar is fundamentally new and non-trivial.} We adopt the worst-case PCRB among all possible real PDFs as the sensing performance metric, and formulate the beamforming optimization problem to minimize the worst-case PCRB subject to a transmit power constraint. This problem is extremely difficult to solve since: \emph{i)} the possible real PDFs lie in a continuous uncertainty set, which leads to \emph{infinitely many constraints}; \emph{ii)} the PCRB is a complex function of each real PDF, which is difficult to express explicitly; \emph{iii)} the transmit beamforming design needs to strike an optimal balance among all possible real PDFs. To resolve these difficulties, we leverage the \emph{second-order Taylor expansion} to obtain a tractable quadratic approximation of each constraint, based on which we further leverage the \emph{S-procedure} to transform the infinite constraints into one single linear matrix inequality (LMI). The problem is then reformulated as a convex semi-definite program (SDP) which can be solved in polynomial time. The proposed solution approaches the optimal solution as the uncertainty radius decreases, due to the increased accuracy in second-order approximation. Numerical results show that our proposed design significantly outperforms non-robust beamforming and achieves close performance to the optimal solution.

\section{System Model}\label{sec_sys}

We consider a MIMO radar system where a multi-antenna BS equipped with $N_{\mathrm{T}}\geq 1$ transmit antennas and $N_{\mathrm{R}}\geq 1$ co-located receive antennas aims to sense the \emph{unknown} and \emph{random} angular location (angle) of a point target denoted by $\theta\in [-\frac{\pi}{2},\frac{\pi}{2})$. Specifically, the BS adopts a uniform linear array (ULA) configuration for both the transmitter and the receiver, and is located at the origin of a two-dimensional (2D) polar coordinate system, as illustrated in Fig. \ref{system model}. The BS sends a sequence of $L>1$ probing signals to estimate the angle based on the echo signals reflected by the target and received back at the BS receive antennas as well as the known prior distribution information of the target.  \looseness=-1

We consider a line-of-sight (LoS) channel between the BS and the target. Let $\alpha=\alpha_{\mathrm{R}}+j\alpha_{\mathrm{I}}\in \mathbb{C}$ denote the overall complex reflection coefficient, which contains the round-trip channel gain and the radar cross-section (RCS) coefficient. Note that $\alpha$ or equivalently $(\alpha_{\mathrm{R}},\alpha_{\mathrm{I}})$ is generally unknown, while its distribution denoted by  $p_{\alpha_{\mathrm{R}},\alpha_{\mathrm{I}}}(\alpha_{\mathrm{R}},\alpha_{\mathrm{I}})$ can be known \emph{a priori} based on target properties.\footnote{We consider a mild condition of $\iint\alpha_{\mathrm{R}}p_{\alpha_{\mathrm{R}},\alpha_{\mathrm{I}}}(\alpha_{\mathrm{R}},\alpha_{\mathrm{I}})d\alpha_{\mathrm{R}}d\alpha_{\mathrm{I}}=\iint\alpha_{\mathrm{I}}p_{\alpha_{\mathrm{R}},\alpha_{\mathrm{I}}}(\alpha_{\mathrm{R}},\alpha_{\mathrm{I}})d\alpha_{\mathrm{R}}d\alpha_{\mathrm{I}}=0$, which holds for various random variables (RVs), including but not limited to all proper RVs.\looseness=-1} Moreover, let $\bm{a}^H(\theta)\in \mathbb{C}^{1\times N_{\mathrm{T}}}$ and $\bm{b}(\theta)\in \mathbb{C}^{N_{\mathrm{R}}\times 1}$ denote the steering vectors of the transmit and receive antennas, respectively, with each element given by $a_p^*(\theta)=e^{\frac{j\pi(N_{\mathrm{T}}-2p+1)\sin\theta}{2}}$, $p=1,...,N_{\mathrm{T}}$, and $b_q(\theta)=e^{\frac{-j\pi(N_{\mathrm{R}}-2q+1)\sin\theta}{2}}$, $q=1,...,N_{\mathrm{R}}$, respectively. The equivalent MIMO channel from the BS transmitter to the BS receiver via target reflection is thus given by $\bm{H}(\alpha,\theta)=\alpha\bm{b}(\theta)\bm{a}^H(\theta)$.
\looseness=-1

Let $\bm{x}_l\in\mathbb{C}^{N_{\mathrm{T}}\times 1}$ denote the baseband equivalent probing signal vector in the $l$-th sample, $l=1,...,L$. The sample covariance matrix is thus given by $\bm{W}=\frac{1}{L}\sum_{l=1}^L\bm{x}_l\bm{x}_l^H$. Let $P$ denote the transmit power budget, which yields $\mathrm{tr}(\bm{W})\leq P$.
The received echo signal vector in each $l$-th sample is given by \looseness=-1
\begin{align}\label{y_l_sensing}
	\bm{y}_l\!=\!\bm{H}(\alpha,\theta)\bm{x}_l\!+\!\bm{n}_l\!=\!\alpha\bm{b}(\theta)\bm{a}^H(\theta)\bm{x}_l\!+\!\bm{n}_l,\ l=1,...,L,
\end{align}
where $\bm{n}_l\sim\mathcal{CN}(\bm{0}, \sigma^2\bm{I}_{N_{\mathrm{R}}})$ denotes the circularly symmetric complex Gaussian (CSCG) noise vector at the BS receiver, with $\sigma^2$ denoting the average noise power. The collection of received signal vectors available for sensing can be expressed as
\begin{equation}
	\bm{Y}\!=\![\bm{y}_1,...,\bm{y}_L ]\!=\!\alpha\bm{b}(\theta)\bm{a}^H(\theta)[ \bm{x}_1 ,...,\bm{x}_L ]\!+\![\bm{n}_1,...,\bm{n}_L ].
\end{equation}

In this paper, we focus on a challenging scenario where the prior distribution information of the angle to be sensed is \emph{imperfect}. Specifically, denote the \emph{imperfect PDF} of $\theta$ known at the BS as $\bar{p}_{\Theta}(\theta)$, and the \emph{real (actual) PDF} of $\theta$ as $p_{\Theta}(\theta)$. $p_{\Theta}(\theta)$ is \emph{unknown} at the BS and modeled as \looseness=-1
\begin{equation}\label{real PDF}
	p_{\Theta}(\theta)=\bar{p}_{\Theta}(\theta)+e(\theta),
\end{equation}
where $e(\theta)\triangleq p_{\Theta}(\theta)-\bar{p}_{\Theta}(\theta)$ is an \emph{unknown} error (or perturbation) function. Note that $e(\theta)$ satisfies $\int e(\theta)d\theta=0$. Furthermore, we assume that the ``energy'' of the error function is bounded, i.e., $\int e^2(\theta)d\theta\le\delta^2$, where $\delta\geq 0$ denotes the uncertainty radius. We assume the value of $\delta$ is \emph{known} to represent the overall level of imperfection in the target's angle PDF.
Consequently, the uncertainty set containing all possible realizations of the real PDF $p_{\Theta}(\theta)$ is also \emph{known} and given by \looseness=-1
\begin{align}\label{P_set}
	\!\!\!\!	\mathcal{P}(\bar{p}_{\Theta}(\theta),\delta)	\triangleq\Big\{p_{\Theta}(\theta)|&p_{\Theta}(\theta)=\bar{p}_{\Theta}(\theta)+e(\theta),\nonumber\\
	& \int e(\theta)d\theta=0, \int e^2(\theta)d\theta\le\delta^2\Big\}.
\end{align}

Note that the imperfect knowledge of the prior distribution information critically affects the transmit signal design (or ``\emph{beamforming}'') and sensing performance. For example, the beamforming design according to the imperfectly known PDF $\bar{p}_{\Theta}(\theta)$ may focus too much power over angles where $\bar{p}_{\Theta}(\theta)$ is large, while the real probability density $p_{\Theta}(\theta)$ is actually small, as illustrated in Fig. \ref{system model}. This will lead to wastage of transmit power and lack of sufficient power focusing over actual highly-probable angles. In this paper, we aim to propose a \emph{robust beamforming design} which optimizes the worst-case performance of MIMO radar sensing among all possible real PDFs in $\mathcal{P}(\bar{p}_{\Theta}(\theta),\delta)$, in order to realize robust sensing. 

To reveal fundamental insights on the effect of such inaccurate prior distribution information with respect to $\theta$, we assume the distribution information for $\alpha$ is accurately known, and $\alpha$ is independent of $\theta$. In the following, we first characterize the worst-case sensing performance, based on which we propose a robust beamforming optimization approach. \looseness=-1

\vspace{-2mm}
\section{Characterization of Worst-Case Sensing Performance via Worst-Case PCRB}\label{sec_WC_PCRB}
\vspace*{-1mm}

\addtolength{\topmargin}{0.03in}

In this section, we aim to characterize the worst-case sensing performance over all possible real PDFs in $\mathcal{P}(\bar{p}_{\Theta}(\theta),\delta)$. Note that the BS only has imperfect PDF information, thus the BS receiver can only estimate the desired parameter based on such imperfect PDF. Since the MSE even for the ideal case with perfect PDF cannot be explicitly expressed, we consider a global lower bound for the MSE with any estimator. Specifically, note that the MSE for sensing exploiting imperfect prior PDF information at the receiver is always lower bounded by the minimum MSE exploiting perfect (real) prior PDF information at the receiver, while the latter is further lower bounded by the PCRB based on the perfect (real) prior PDF information, which has an explicit expression and is tight in moderate-to-high SNR regimes \cite{van1968detection}. Motivated by this, we adopt the PCRB based on real prior PDF as the sensing performance metric, which is increasingly tight with respect to the MSE with imperfect prior PDF as the uncertainty radius $\delta$ decreases in moderate-to-high SNR regimes.\footnote{It is worth noting that there is also another type of mis-specified PCRB (MPCRB) \cite{fortunati2017performance} which may account for the mismatch between the assumed and true models. However, MPCRB has a limited application range (e.g., for a certain class of estimators) and involves even more complex operations over the real PDF. The extension of this work along the line of MPCRB is left as our future work.} \looseness=-1

Let $\bm{\zeta}=[\theta, \alpha_{\mathrm{R}}, \alpha_{\mathrm{I}}]^T$ denote the collection of all unknown (real) parameters, which need to be jointly estimated to obtain an accurate estimate of $\theta$. The real PDF of $\bm{\zeta}$ is given by $p_Z(\bm{\zeta})=p_{\Theta}(\theta)p_{\alpha_{\mathrm{R}},\alpha_{\mathrm{I}}}(\alpha_{\mathrm{R}},\alpha_{\mathrm{I}})$. The posterior Fisher information matrix (PFIM) for sensing $\bm{\zeta}$ is given by $\bm{F}=\bm{F}_{\mathrm{O}}+\bm{F}_{\mathrm{P}}$, where $\bm{F}_{\mathrm{O}}\in\mathbb{R}^{3\times 3}$ denotes the PFIM from the observations in $\bm{Y}$, and $\bm{F}_{\mathrm{P}}\in\mathbb{R}^{3\times 3}$ denotes the PFIM from the prior distribution information in $p_Z(\bm{\zeta})$. $\bm{F}_{\mathrm{O}}$ is given by \looseness=-1
\begin{align}\label{F_O}
	\!\!\!\!\bm{F}_{\mathrm{O}}\!\!=\!\mathbb{E}_{\bm{Y}\!\!,\bm{\zeta} }\!\bigg[\!\frac{\partial \ln(f(\bm{Y}|\bm{\zeta}))}{\partial \bm{\zeta} }\!\bigg(\!\!\frac{\partial \ln(f(\bm{Y}|\bm{\zeta}))}{\partial \bm{\zeta} }\!\bigg)^{\!\!H}\bigg]\!\!\!=\!\!\!
	\begin{bmatrix}
		F^{\theta\theta}_{\mathrm{O}} &\!\!\!\!\!\! \bm{F}^{\theta\alpha}_{\mathrm{O}}\\
		\bm{F}^{{\theta\alpha}^H}_{\mathrm{O}} & \!\!\!\!\!\!\bm{F}^{\alpha\alpha}_{\mathrm{O}}
	\end{bmatrix}\!\!,
\end{align}
where $f(\bm{Y}|\bm{\zeta})$ denotes the conditional PDF of $\bm{Y}$ given $\bm{\zeta}$. Note that $\bm{F}_{\mathrm{O}}$ is determined by the real PDF $p_{\Theta}(\theta)=\bar{p}_{\Theta}(\theta)+e(\theta)$. Specifically, define $\bm{M}(\theta)\triangleq\bm{b}(\theta)\bm{a}^H(\theta)$, 
$\Dot{\bm{M}}(\theta)\triangleq\frac{\partial\bm{M}(\theta)}{\partial\theta}$, and $\gamma\triangleq\iint(\alpha_{\mathrm{R}}^2+\alpha_{\mathrm{I}}^2)p_{\alpha_{\mathrm{R}},\alpha_{\mathrm{I}}}(\alpha_{\mathrm{R}},\alpha_{\mathrm{I}})d\alpha_{\mathrm{R}}d\alpha_{\mathrm{I}}$. Each block in $\bm{F}_{\mathrm{O}}$ can be derived as
\begin{align}
	F^{\theta\theta}_{\mathrm{O}}
	&=\frac{2L\gamma}{\sigma^2}\mathrm{tr}\left(\left(\int\Dot{\bm{M}}^H(\theta)\Dot{\bm{M}}(\theta){p}_{\Theta}(\theta)d\theta\right)\bm{W}\right),\\
	\bm{F}^{\theta\alpha}_{\mathrm{O}}
	&=\bm{0},\\
	\bm{F}^{\alpha\alpha}_{\mathrm{O}}
	&=\frac{2L}{\sigma^2}\mathrm{tr}\left(\left(\int\bm{M}^H(\theta)\bm{M}(\theta){p}_{\Theta}(\theta))d\theta\right)\bm{W}\right)\bm{I}_2.
\end{align}
Moreover, $\bm{F}_\mathrm{P}$ can be derived as 
\begin{align}\label{F_P}
	\!\!\!\!	\bm{F}_{\mathrm{P}}\!\!=\!\!\mathbb{E}_{\bm{\zeta} }\!\left[\frac{\partial \ln(p_Z(\bm{\zeta} ))}{\partial \bm{\zeta} }\left(\frac{\partial \ln(p_Z(\bm{\zeta} ))}{\partial \bm{\zeta} }\right)^H\right] 
	\!\!=\!\!\begin{bmatrix}
		F_{\mathrm{P}}^{\theta\theta} & \!\!\!\! \bm{0}\\
		\bm{0} &\!\!\!\! \bm{F}^{\alpha\alpha}_{\mathrm{P}}
	\end{bmatrix}\!\!,
\end{align}
where  $[\bm{F}^{\alpha\alpha}_{\mathrm{P}}]_{m,n}\!\!=\!\mathbb{E}_{\alpha_{\mathrm{R}},\alpha_{\mathrm{I}}}\!\bigg[\frac{\partial \ln(p_{\alpha_{\mathrm{R}},\alpha_{\mathrm{I}}}(\alpha_{\mathrm{R}},\alpha_{\mathrm{I}}))}{\partial \bar{\alpha}_m }\frac{\partial \ln(p_{\alpha_{\mathrm{R}},\alpha_{\mathrm{I}}}(\alpha_{\mathrm{R}},\alpha_{\mathrm{I}}))}{\partial \bar{\alpha}_n }\!\bigg]$ with
$\bar{\alpha}_1=\alpha_{\mathrm{R}},\bar{\alpha}_2=\alpha_{\mathrm{I}}$; $F_{\mathrm{P}}^{\theta\theta}$ is given by
\begin{align}\label{J_P}
	\!\!\!  F_{\mathrm{P}}^{\theta\theta}\!\!=\!\mathbb{E}_\theta\Big[\Big(\frac{\partial\ln p_{\Theta}(\theta)}{\partial\theta}\Big)^{\!\!2}\Big]
	\!\!=\!\!\!\int \!\! \frac{\big(\frac{\partial p_{\Theta}(\theta)}{\partial\theta}\big)^2}{{p}_{\Theta}(\theta)}d\theta.
\end{align}
Based on the above, the PFIM for $\bm{\zeta}$ is given by
\begin{equation}
	\bm{F}\!=\bm{F}_{\mathrm{O}}+\bm{F}_{\mathrm{P}}=\!\!\begin{bmatrix}
		F^{\theta\theta}_{\mathrm{O}}+F_{\mathrm{P}}^{\theta\theta} &\!\!\!\!
		\bm{0}\\
		\bm{0} &\!\!\!\! \bm{F}^{\alpha\alpha}_{\mathrm{O}}+\bm{F}^{\alpha\alpha}_{\mathrm{P}}
	\end{bmatrix}.
\end{equation}

The overall PCRB matrix for the MSE matrix in estimating $\bm{\zeta}$ is given by $\bm{F}^{-1}$. The PCRB for the MSE in estimating the desired sensing parameter $\theta$ is given by
\begin{align}
	&\mathrm{PCRB}_{\theta}\\
	=&\frac{1}{\frac{2L\gamma}{\sigma^2}\mathrm{tr}\left(\left(\!\int\!\!\Dot{\bm{M}}^H\!(\theta)\Dot{\bm{M}}(\theta)p_{\Theta}(\theta)d\theta\right)\bm{W}\!\right)+\!\!\!\int \!\! \frac{\big(\frac{\partial p_{\Theta}(\theta)}{\partial\theta}\big)^2}{p_{\Theta}(\theta)}d\theta\!}.\nonumber
\end{align}
\textls[-10]{Notice that each $\mathrm{PCRB}_{\theta}$ is inversely proportional to a linear function of the sample covariance matrix $\bm{W}$, where both the ``slope'' and added constant are critically determined by the real prior PDF $p_\Theta(\theta)$. The worst-case PCRB over all possible real PDFs $p_{\Theta}(\theta)$'s in the uncertainty set $\mathcal{P}(\bar{p}_{\Theta}(\theta),\delta)$ is thus given by} \looseness=-1
\begin{align}\label{worst case PCRB}
	\mathrm{PCRB}_{\theta}^{\mathrm{worst}}\triangleq\underset{p_{\Theta}(\theta)\in\mathcal{P}(\bar{p}_{\Theta}(\theta),\delta)}{\max}\mathrm{PCRB}_{\theta}.
\end{align}
Note that $\mathrm{PCRB}_{\theta}^{\mathrm{worst}}$ serves as a tight lower bound of the worst-case MSE among all possible real PDFs in moderate-to-high SNR regimes when the uncertainty radius $\delta$ is small.

In the following, we study the optimization of $\bm{W}$ in a robust manner with $\mathrm{PCRB}_{\theta}^{\mathrm{worst}}$ as the sensing performance metric.

\vspace{-1mm}
\section{Problem Formulation}

We aim to optimize the transmit sample covariance matrix $\bm{W}$ to minimize the worst-case PCRB over all possible real prior PDFs in the uncertainty set $\mathcal{P}(\bar{p}_\Theta(\theta),\delta)$ defined in (\ref{P_set}), based on knowledge of the imperfect prior PDF $\bar{p}_\Theta(\theta)$ and the uncertainty radius $\delta$ which bounds the error between the real and imperfect prior PDFs. By introducing an auxiliary variable $t$, the optimization problem is formulated as:
\begin{align}
	\!\!\mbox{(P1)} \underset{t,\bm{W}\succeq \bm{0}}{\max} & t \label{P1_obj}\\
	\mathrm{s.t.}\  &\frac{2L\gamma}{\sigma^2}\mathrm{tr}\left(\left(\int\Dot{\bm{M}}^H(\theta)\Dot{\bm{M}}(\theta)p_\Theta(\theta)d\theta\right)\bm{W}\right)\nonumber \\
	& \!+\!\int \!\! \frac{\big(\frac{\partial p_\Theta(\theta)}{\partial\theta}\big)^2}{p_{\Theta}(\theta)}d\theta\ge t,\ \forall p_\Theta(\theta)\!\in\! \mathcal{P}(\bar{p}_{\Theta}(\theta),\delta)    \label{P1_constraint1}\\
	& \mathrm{tr}(\bm{W}) \leq P. \label{P1_power constraint}
\end{align}

Problem (P1) is highly non-trivial due to the following reasons. \emph{Firstly}, due to the continuity of the uncertainty set $\mathcal{P}(\bar{p}_{\Theta}(\theta),\delta)$ and the continuous domain of $\theta$, there are \emph{infinite} possible real PDFs $p_\Theta(\theta)$'s, which corresponds to \emph{infinite constraints} in (\ref{P1_constraint1}). \emph{Secondly}, $\mathcal{P}(\bar{p}_{\Theta}(\theta),\delta)$ and each possible $p_\Theta(\theta)$ therein are only bounded around $\bar{p}_\Theta(\theta)$ via $\delta$, while the explicit expressions are not available. This makes it extremely difficult to express each constraint in (\ref{P1_constraint1}) which is particularly challenging due to the involvement of complex operations on each $p_\Theta(\theta)$ such as differentiation and division. \emph{Finally}, different real PDF $\bar{p}_{\Theta}(\theta)$ leads to different constant terms in (\ref{P1_constraint1}). Hence, one design of $\bm{W}$ needs to judiciously balance among all the possible $p_\Theta(\theta)$'s in the uncertainty set $\mathcal{P}(\bar{p}_{\Theta}(\theta),\delta)$, which makes the studied robust design fundamentally different from the design in prior works with perfect prior PDF.

In the following, we overcome these challenges via equivalent transformation and effective approximation.

\addtolength{\topmargin}{0.02in}

\vspace{-1mm}
\section{Proposed Solution to (P1)} \label{sec:s-procedure}
\vspace{-1mm}

In this section, we present our proposed solution to the robust beamforming optimization problem in (P1). Specifically, we first approximate each constraint in (\ref{P1_constraint1}) as a tractable quadratic function of the real PDF and equivalently the error function via \emph{second-order Taylor expansion}. Then, we equivalently transform the infinitely many constraints in (\ref{P1_constraint1}) into a single LMI via \emph{S-procedure}, by exploiting the bounded nature of the error function. This thus transforms (P1) into a convex SDP with a small finite number of constraints.\looseness=-1

\subsection{Second-Order Approximation of Each Constraint in (\ref{P1_constraint1})}\label{sec:s-procedure A}

In this subsection, we approximate each constraint in (\ref{P1_constraint1}) as a tractable form, for which the key challenge lies in the complex differentiation and division involved in $\int \!\! \frac{\big(\frac{\partial p_{\Theta}(\theta)}{\partial\theta}\big)^2}{{p}_{\Theta}(\theta)}d\theta$. For ease of exposition, we expand $p_{\Theta}(\theta)$ as $\bar{p}_{\Theta}(\theta)+e(\theta)$ in the following. To address this challenge, we propose to apply second-order Taylor expansion on $\frac{1}{\bar{p}_{\Theta}(\theta)+e(\theta)}$ around $e(\theta)=0$ since the error probability density is typically much smaller than the real/imperfect probability density, which is given by
\begin{equation}\label{Taylor second expansion}
	\frac{1}{\bar{p}_{\Theta}(\theta) + e(\theta)} \approx \frac{1}{\bar{p}_{\Theta}(\theta)} - \frac{e(\theta)}{\bar{p}_{\Theta}^2(\theta)} + \frac{e^2(\theta)}{\bar{p}_{\Theta}^3(\theta)}.
\end{equation}
By substituting (\ref{Taylor second expansion}) into (\ref{P1_constraint1}), we further have
\begin{align}\label{F_P approx}
	&\!\!\!\!\!\int \!\! \frac{\big(\frac{\partial p_{\Theta}(\theta)}{\partial\theta}\!\big)^2}{{p}_{\Theta}(\theta)}d\theta
	\!\approx\!\! \!\int\!\!\Big(
	\frac{(\frac{\partial\bar{p}_{\Theta}(\theta)}{\partial\theta})^2}{\bar{p}_{\Theta}(\theta)}
	\!+\!\frac{2\frac{\partial\bar{p}_{\Theta}(\theta)}{\partial\theta}\frac{\partial e(\theta)}{\partial\theta}}{\bar{p}_{\Theta}(\theta)}
	\!-\!\frac{(\frac{\partial\bar{p}_{\Theta}(\theta)}{\partial\theta})^2e(\theta)}{\bar{p}_{\Theta}^2(\theta)}
	\nonumber\\
	&+\!\frac{(\frac{\partial\bar{p}_{\Theta}(\theta)}{\partial\theta})^2e^2(\theta)}{\bar{p}_{\Theta}^3(\theta)}
	\!-\!\frac{2\frac{\partial\bar{p}_{\Theta}(\theta)}{\partial\theta}\frac{\partial e(\theta)}{\partial\theta}e(\theta)}{\bar{p}_{\Theta}^2(\theta)}
	\!+\!\frac{(\frac{\partial e(\theta)}{\partial\theta})^2}{\bar{p}_{\Theta}(\theta)}
	\Big)d\theta.
\end{align}

Note that (\ref{F_P approx}) is now in a quadratic form with respect to $e(\theta)$ as well as its derivative. To make it more tractable, we propose to quantize the feasible range of $\theta\in[-\frac{\pi}{2},\frac{\pi}{2})$ into $N>1$ discrete grid points with uniform spacing $\Delta\theta$. Let $\bar{\bm{p}}\in\mathbb{R}^{N\times 1}$ and $\bm{e}\in\mathbb{R}^{N\times 1}$ denote the discretized versions of the imperfect prior PDF $\bar{p}_{\Theta}(\theta)$ and the error function $e(\theta)$, respectively. After discretization, the differentiation operation $\frac{\partial}{\partial\theta}$ is conducted via multiplication with a difference matrix $\bm{D}\in\mathbb{R}^{N\times N}$, while the integration operation is conducted via the Riemann sum. Define $\bm{\Sigma}_1=\mathrm{diag}(\bar{\bm{p}})^{-1}$, $\bm{\Sigma}_2=\mathrm{diag}(\bar{\bm{p}}\odot\bar{\bm{p}})^{-1}$, $\bm{\Sigma}_3=\mathrm{diag}(\bar{\bm{p}}\odot\bar{\bm{p}}\odot\bar{\bm{p}})^{-1}$, and $\bm{V}_0=\mathrm{diag}(\bm{D}\bar{\bm{p}})$, where $\odot$ denotes the  Hadamard product. By further defining $C_{\mathrm{P}}\!\!=\!\!(\bm{D}\bar{\bm{p}})^T\bm{\Sigma}_1(\bm{D}\bar{\bm{p}})\Delta\theta$, $\bm{Q}_{\mathrm{P}} \!\!=\! \!( \bm{D}^T \bm{\Sigma}_1 \bm{D} \!-\! (\bm{\Sigma}_2 \bm{V}_0 \bm{D} \!+\! \bm{D}^T\bm{V}_0\bm{\Sigma}_2) \!+\! \bm{\Sigma}_3 \bm{V}_0^2 ) \Delta\theta\!\!\in\!\!\mathbb{R}^{N\times N}$, and $\bm{f}_{\mathrm{P}}\!=\!( 2 \bm{D}^T \bm{\Sigma}_1 (\bm{D}\bar{\bm{p}}) - \bm{\Sigma}_2 \bm{V}_0 (\bm{D}\bar{\bm{p}})) \Delta\theta\!\in\!\mathbb{R}^{N\times 1}$ which are constants determined by the imperfect known PDF, we have
\begin{equation}
	\int \!\! \frac{\big(\frac{\partial p_{\Theta}(\theta)}{\partial\theta}\big)^2}{{p}_{\Theta}(\theta)}d\theta \approx \bm{e}^T \bm{Q}_{\mathrm{P}} \bm{e} + \bm{f}_{\mathrm{P}}^T \bm{e} + C_{\mathrm{P}},
\end{equation}
which is a \emph{quadratic function} of the discretized error function. 

Similarly, define $\bm{g}(\bm{W})\in\mathbb{R}^{N\times 1}$ with $g_n(\bm{W})=\frac{2L\gamma\Delta\theta}{\sigma^2}\mathrm{tr}(\Dot{\bm{M}}^H(\theta_n)\Dot{\bm{M}}(\theta_n)\bm{W}),\forall n$. We then have
\begin{align}\label{J_D}
	&\frac{2L\gamma}{\sigma^2}\mathrm{tr}\left(\left(\int\Dot{\bm{M}}^H(\theta)\Dot{\bm{M}}(\theta)p_\Theta(\theta)d\theta\right)\bm{W}\right) \nonumber\\
	\approx& \bar{\bm{p}}^T \bm{g}(\bm{W}) + \bm{e}^T \bm{g}(\bm{W}),
\end{align}
where the approximation is purely due to discretization.

By further noting that the discretized versions of $\int e(\theta)d\theta=0$ and $\int e^2(\theta)d\theta\le\delta^2$ are given by $\bm{1}^T\bm{e}=0$ and $\|\bm{e}\|^2\leq \frac{\delta^2}{\Delta\theta}$, respectively, Problem (P1) is approximated as
\begin{align}
	\!\!	\mbox{(P2)}\ \underset{t,\bm{W}\succeq \bm{0}}{\max}\ & t \label{P2_obj}\\
	\mathrm{s.t.}\ &\bm{e}^T \bm{Q}_{\mathrm{P}} \bm{e}+(\bm{f}_{\mathrm{P}}+\bm{g}(\bm{W}))^T\bm{e}+C_{\mathrm{P}}+\bar{\bm{p}}^T\bm{g}(\bm{W})\geq t, \nonumber\\
	&\qquad\qquad \forall \bm{e}\in \bigg\{\bm{e}|\bm{1}^T\bm{e}=0,\|\bm{e}\|^2\leq \frac{\delta^2}{\Delta\theta}\bigg\}\!\!\!\!   \label{P2_constraint1}\\
	& \mathrm{tr}(\bm{W}) \leq P.
\end{align}
It is worth noting that (P2) is equivalent to (P1) when the error function is sufficiently small (which holds when $\delta$ is sufficiently small) and the discretization granularity $\Delta \theta$ is sufficiently small. Moreover, each constraint in (\ref{P2_constraint1}) (the original (\ref{P1_constraint1})) is in a simple quadratic form over the discretized error function $\bm{e}$, which is further bounded in linear and quadratic constraints. However, there are still \emph{infinitely many constraints} in (\ref{P2_constraint1}) due to the continuous feasible set of $\bm{e}$, as will be addressed below.

\subsection{Equivalent Transformation of Infinite Constraints via S-Procedure}

In this subsection, we aim to leverage the bounded feature of $\bm{e}$ to transform the infinitely many constraints in (\ref{P2_constraint1}) into an equivalent constraint. To this end, we first define an auxiliary vector $\bm{u}\in\mathbb{R}^{(N-1)\times1}$ with $\bm{e}=\bm{B}\bm{u}$, where $\bm{B}\in\mathbb{R}^{N\times (N-1)}$ denotes an orthogonal basis matrix for the null space of $\bm{1}^T$. Then, (\ref{P2_constraint1}) is equivalently expressed as
\begin{align} \label{c3}
	&t-(\bm{u}^T \bm{B}^T\bm{Q}_{\mathrm{P}}\bm{B} \bm{u} + \bm{B}^T(\bm{f}_{\mathrm{P}}+\bm{g}(\bm{W}))^T \bm{u} + C_{\mathrm{P}}+\bar{\bm{p}}^T\bm{g}(\bm{W}))\nonumber\\ &\leq 0,\quad 
	\forall \bm{u}\in \left\{\bm{u}|\|\bm{u}\|^2-\frac{\delta^2}{\Delta\theta} \le 0\right\}.
\end{align}

Note that the equivalent new constraint in (\ref{c3}) aims to make sure a quadratic inequality constraint with respect to $\bm{u}$ is satisfied for all $\bm{u}$'s that satisfy another quadratic inequality constraint, which can be guaranteed via the S-procedure. To this end, we present the following lemma. 
\begin{lemma}\label{lemma1}
	\emph{S-procedure \cite{luo2004multivariate}:} Define quadratic functions $f_i(\bm{x})=\bm{x}^H\bm{T}_i\bm{x}+2\mathfrak{Re}\{\bm{b}_i^H\bm{x}\}+c_i$, $i=1,2$, where $\bm{T}_i\in\mathbb{C}^{N\times N}$, $\bm{b}_i\in\mathbb{C}^{N\times 1}$, and $c_i\in\mathbb{R}$. $f_1(\bm{x})\leq 0$ guarantees $f_2(\bm{x})\leq 0$ if and only if there exists $\lambda\geq 0$ such that
	\begin{equation}
		\lambda\begin{bmatrix}
			\bm{T}_1 & \bm{b}_1\\
			\bm{b}_1^H & c_1
		\end{bmatrix}-\begin{bmatrix}
			\bm{T}_2 & \bm{b}_2\\
			\bm{b}_2^H & c_2
		\end{bmatrix}\succeq\bm{0}.
	\end{equation}
\end{lemma}

By applying the S-procedure in Lemma \ref{lemma1}, Problem (P2) is equivalently transformed as the following problem:  \looseness=-1
\begin{align}
	\mbox{(P3)}\, \underset{t,\lambda\geq 0,\bm{W}\succeq \bm{0}}{\max}\, & t \label{P3_obj}\\
	\mbox{s.t.}\ & \begin{bmatrix}
		\bm{\Phi}_{11}(\lambda) & \bm{\Phi}_{12}(\bm{W})\\
		\bm{\Phi}_{12}^H(\bm{W}) & \Phi_{22}(\bm{W}, t, \lambda)
	\end{bmatrix}\succeq\bm{0} \label{P3_constraint1}\\
	& \mathrm{tr}(\bm{W}) \leq P,
\end{align}
where $\bm{\Phi}_{11}(\lambda) = \lambda \bm{I}_{N-1} + \bm{B}^T \bm{Q}_{\mathrm{P}} \bm{B}$, $\bm{\Phi}_{12}(\bm{W}) = \frac{1}{2} \bm{B}^T ( \bm{f}_{\mathrm{P}} + \bm{g}(\bm{W}))$, and $\Phi_{22}(\bm{W}, t, \lambda) = C_{\mathrm{P}} + \bar{\bm{p}}^T \bm{g}(\bm{W}) - t - \lambda \frac{\delta^2}{\Delta\theta}$. Note that the infinitely many constraints in (P2) are now equivalently transformed into an LMI in (\ref{P3_constraint1}). (P3) is a convex SDP, for which the optimal solution can be obtained via the interior-point method \cite{CVX} or existing software, e.g., CVX \cite{cvxtool}. 

Due to the equivalence between (P2) and (P3), the obtained optimal solution to (P3) approaches the optimal solution to (P1) as the uncertainty radius $\delta$ and discretization granularity $\Delta\theta$ decreases. The complexity for our proposed design can be shown to be $\mathcal{O}(N_{\mathrm{T}}^7+N_{\mathrm{T}}^5N^2+N_{\mathrm{T}}^3N^3+N_{\mathrm{T}}^2N_{\mathrm{R}}N)$ \cite{CVX}, which is \emph{polynomial}, despite the infinitely many constraints in (P1). \looseness=-1

\section{Numerical Results}

In this section, we provide numerical results to evaluate the performance of the proposed robust beamforming design. We set $N_{\mathrm{T}}\!=\!10$, $N_{\mathrm{R}}\!=\!12$, $P\!=\!30$ dBm, $\sigma^2\!=\!-90$ dBm, $\Delta\theta\!=\!0.005$ rad, and $\alpha\!\sim\!\mathcal{CN}(0,2\!\times\! 10^{-14})$. We consider a Gaussian mixture model for the imperfect known PDF, where $\bar{p}_{\Theta}(\theta)\!=\!\sum_{k=1}^K\frac{p_{\mathrm{N},k}}{\sqrt{2\pi}\sigma_{\mathrm{N},k}}e^{-\frac{(\theta-\theta_{\mathrm{N},k})^2}{2\sigma_{\mathrm{N},k}^2}}$ with $K\!=\!2$; $\theta_{\mathrm{N},1}\!=\!-0.7$, $\theta_{\mathrm{N},2}\!=\!0.7$; $\sigma_{\mathrm{N},1}^2\!=\!10^{-3}$, $\sigma_{\mathrm{N},2}^2\!=\!10^{-2.8}$; and $p_{\mathrm{N},1}\!=\!0.6$, $p_{\mathrm{N},2}\!=\!0.4$. We further consider the following benchmark schemes.
\looseness=-1
\begin{itemize}
	\item \textbf{Benchmark Scheme 1: Non-robust beamforming.} This scheme optimizes $\bm{W}$ to minimize the PCRB based on the imperfect PDF $\bar{p}_{\Theta}(\theta)$, which is a convex problem.
	\item \textbf{Benchmark Scheme 2: Enumeration-based robust beamforming.} This scheme uses $M>1$ samples to approximate the continuous set $\mathcal{P}(\bar{p}_{\Theta}(\theta),\delta)$, and enumerate all of them in $M$ constraints to approximate (\ref{P1_constraint1}). Then, (P1) can be solved via the interior-point method with complexity $\mathcal{O}(N_{\mathrm{T}}^7+N_{\mathrm{T}}^5M^2+N_{\mathrm{T}}^3M^3)$. Note that the accuracy of scheme increases as $M$ increases, which leads to prohibitive complexity.  \looseness=-1
\end{itemize}

We evaluate the worst-case PCRB of the proposed and benchmark schemes over $1000$ realizations of the real PDF under a Gaussian mixture model with $p_{\Theta}(\theta)=\sum_{i=1}^3\frac{p_{\mathrm{T},i}}{\sqrt{2\pi}\sigma_{\mathrm{T},i}}e^{-\frac{(\theta-\theta_{\mathrm{T},i})^2}{2\sigma_{\mathrm{T},i}^2}}$, where each realization is randomly generated by varying the weights, means, and variances of the three Gaussian components, bounded by the design uncertainty radius $\delta$. We assume these realizations are exactly the samples used in a genie-aided Benchmark Scheme 2, thus it corresponds to a \emph{genie-aided lower bound} of the worst-case PCRB with extra information about the uncertainty set.

\begin{figure}[t]
	\centering
	\includegraphics[width=0.48\textwidth]{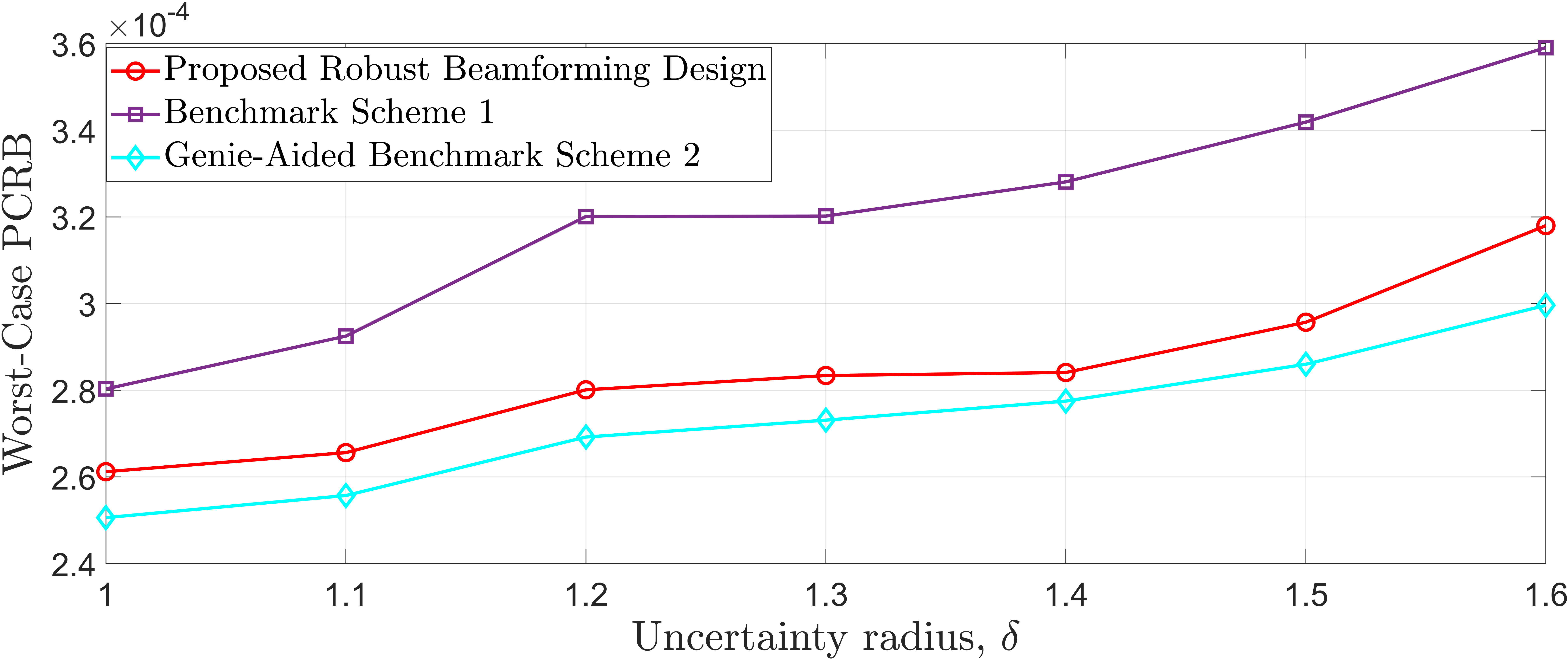}
	\vspace{-3mm}
	\caption{Worst-case PCRB versus uncertainty radius $\delta$ for different schemes.}
	\label{PCRB_VS_delta}
	\vspace{0mm}\end{figure}

Fig. \ref{PCRB_VS_delta} shows the worst-case PCRB achieved by different schemes versus the uncertainty radius $\delta$. It is observed that the worst-case PCRB for all schemes increases with $\delta$, since a larger uncertainty radius implies a larger set of admissible real PDFs, automatically degrading the worst-case sensing performance. Moreover, both the proposed scheme and Benchmark Scheme 2 significantly outperform Benchmark Scheme 1, due to the judicious robust beamforming design considering prior PDF imperfection. Furthermore, the proposed scheme performs closely to the genie-aided Benchmark Scheme 2, where all the real PDFs in the evaluation set are perfectly known and fully used. This thus validates the tightness of our proposed approximations and effectiveness of the S-procedure.

On the other hand, it is worth noting that our proposed scheme consumes significantly lower computational time compared with Benchmark Scheme 2 thanks to the S-procedure. Under $\delta=1.2$ and $\delta=1.6$, Benchmark Scheme 2 with $M=1000$ samples takes $14.846$ seconds (s) and $14.761$ s, while our proposed scheme only takes $5.987$ s and $5.968$ s, with $59.67\%$ and $59.57\%$ time reduction, respectively.

\begin{figure}[t]
	\vspace{2mm}
	\centering
	\includegraphics[width=0.48\textwidth]{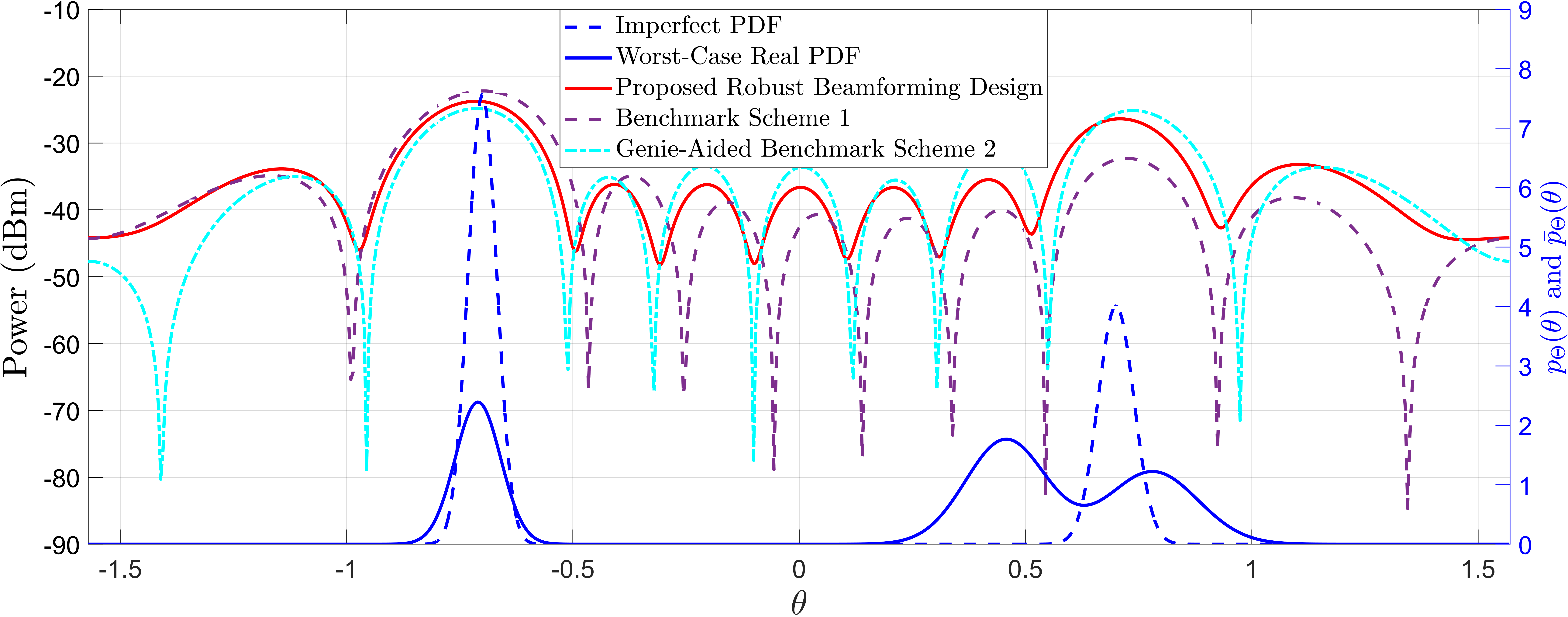}
	\vspace{-3mm}
	\caption{Radiated power patterns of different schemes under $\delta=1.6$.}
	\label{radiated power pattern}
	\vspace{0mm}
\end{figure} 

Finally, Fig. \ref{radiated power pattern} illustrates the radiated power patterns of the proposed and benchmark schemes at $40$ m under $\delta=1.6$, where the imperfect PDF $\bar{p}_{\Theta}(\theta)$ and the worst-case real PDF $p_{\Theta}(\theta)$ for the proposed scheme are also shown. It is observed that Benchmark Scheme 1 allocates power strictly according to the imperfect PDF by concentrating high power only over the two peaks. In contrast, the proposed scheme generates a power pattern similar to that of the genie-aided Benchmark Scheme 2. Specifically, these two schemes yield a more ``flattened'' power pattern which well-covers potential highly-probable angles surrounding the two peaks in the known PDF, to combat the possible PDF imperfection via proactive robust beamforming.

\section{Conclusions}

This paper studied a MIMO radar system where the unknown and random angle parameter of a point target needs to be sensed based on target-reflected echo signals and prior distribution information. We considered a challenging yet practical case where such information is imperfect, while the actual real PDF is a perturbed version of the imperfect known PDF bounded by a given uncertainty radius. To take into account such PDF imperfection, we advocated a robust beamforming design framework where the beamforming was optimized to minimize the worst-case PCRB among all possible real PDFs in the uncertainty set. Due to the continuity of the uncertainty set as well as the continuous (angle) domain of the PDF, this problem is highly non-trivial with infinitely many challenging constraints corresponding to infinitely many possible real PDFs, each involving a complex function of one real PDF. To tackle this problem, we proposed a second-order Taylor expansion based approximation of each constraint, and further applied the S-procedure technique to equivalently transform the infinite constraints into one LMI, making the resulting problem a convex SDP solvable with polynomial complexity. It was shown via numerical results that our proposed robust beamforming achieves superior performance compared with its non-robust counterpart, and also perform closely to the globally optimal solution. \looseness=-1

\newpage
\newpage
\bibliographystyle{IEEEtran}
\bibliography{reference}
\end{document}